\documentstyle[preprint,aps,epsf,axodraw,12pt]{revtex}
\begin{document}
\tightenlines
\title{Trace and chiral anomalies in QED and their
underlying theory interpretation}
\author{Ji-Feng Yang}
\address{Department of Physics, East China Normal University,
Shanghai 200062, China}
\maketitle
\begin{abstract}
Parametrizing the possible underlying theory or new physics'
decoupling effects in the most general way we reexamined the
validity of canonical trace relation and chiral symmetry in
certain one-loop two-point functions. The anomalies and the
relation between chiral and trace Ward identities are investigated
and interpreted in the perspective of a complete underlying theory
or new physics instead of regularization, with some new phenomena
found as by-products.
\end{abstract}

\newpage
\section{Introduction}
It has been a standard point of view that quantum field theories
for particle physics and others are just effective theories (EFT)
for describing 'low energy' physics in a less sophisticated way
due to UV divergences as crucial underlying short distance
information is lost\cite{Weinbergbook}. However, these
shortcomings can be consistently removed though the
renormalization program that is rigorously established at least in
perturbation theory\cite{BPHZ}. But there arises a new problem
with the renormalization that some classical or canonical
symmetries\footnote{Here and in the following, the symmetry refers
to any classical relations. For example, there could be mass term
that softly violate the exact chiral and scale symmetries} are
violated. As chiral anomaly have direct physical and topological
connections\cite{PCAC} and similarly for trace
anomaly\cite{Crew,PCDC}, such anomalies are often termed as
quantum mechanical violation of classical symmetries. That is, the
quantization procedure is incompatible with such symmetries.

However, in 't Hooft's proposal an alternative view of chiral
anomaly is advanced, the chiral anomaly could arise from the
decoupling of heavy fermions\cite{Hooft}, hence one could alter
the anomaly through adding or removing matter contents. Thus
anomaly is more closely related with dynamical contents, or the
quantum mechanical interpretation of anomaly is not quite
concrete. In other words, anomaly could be replaced by 'something'
normal in a complete underlying theory or framework, like the
long-pursued Theory of Everything. But before this complete theory
is firmly established and profoundly tested, we must work with the
effective theories and parametrize our ignorance somehow. In most
literature the underlying structures (or new physics, in most
authors' terminology) are simply parametrized in a specific
regularization scheme and it is tacitly assumed that the
conclusions are regularization scheme independent. However, such
simplified and regularization-based analysis might not be quite
pertinent or trustworthy, sometimes even misleading\cite{Burgess}.
In our opinion, one should seek as general as possible
parametrization for the underlying or new physics in order to
avoid the limitations associated with a specific regularization.
We emphasize that it is the {\em necessity} of introducing of
regularization but not the form of a regularization imply that
there must be underlying structures and we must parametrize their
existence properly.

In this report, we follow and elaborate on the EFT philosophy (and
the principle implied in 't Hooft's proposal) to reexamine the
chiral and trace Ward identities in the two point functions
involving two axial current(scalar)s. We will calculate the one
loop amplitudes in two approaches: (1) dimensional regularization;
(2) a general differential equation approach. The rationality for
the latter will be given in next section. Here we mention that
this is to try to parametrize all 'possible' underlying theory's
influences.

This paper is organized in the following way: In Sec. II we review
the general considerations concerning the EFT philosophy and the
rationale for the differential equation approach for computing the
loop diagrams. In Sec. III the approach is applied to the two and
three point functions. In Sec. IV we discuss the trace and chiral
Ward identities parametrized by the method introduced in Sec. II.
Then in Sec. V we propose a normal interpretation of any
anomalies. The paper is summarized in Sec. VI.
\section{General considerations from underlying theory scenario}
The EFT philosophy should carry at least two different technical
aspects: (1) the interaction and matter content are not complete
and (2) the theory's quantum structures are not well defined in
the full spectrum. In principle one could imagine that the second
is just a special case of the first one in some sense. But no
matter what kind of underlying theory(new physics) are to be
found, there must be some missing parameters or constants that are
necessary to make the EFT quantities well defined. Then all
phenomena are well defined and normally explicable, that is, no
room for anomaly and ill-definedness. In this sense, an anomaly is
'canonical' in the complete underlying theory of everything
(CUTE). On the other hand, in the 'low energy' ranges
corresponding to EFT's, the underlying degrees become formally
decoupled. But their contributions might induce composite EFT
operators, similar to the well-known Wilson OPE\cite{wilson} and
decoupling of heavy quark fields\cite{Witten}. It is the
appearance of these operators that become 'anomalies' in EFT, as
they are not 'canonically' defined in the EFT formulations, but
come from the decoupling limit of the underlying modes, or new
physics.
\subsection{EFT vs. CUTE: general formulation}
To proceed in a formal way, we assume that it is still appropriate
to use an abstract Hilbert space for CUTE where operators,
coupling constants and state vectors are necessary theoretical
components. We need not speculate about what they look like, we
only need their existence to fill in the gap in the UV ends of EFT
spectra. In other words, we regularize the EFT with the correct
CUTE structures instead of an arbitrary regularization schemes.
Then we could formally define the EFT generating functional (path
integral) in CUTE, and the associated Green functions or vertex
functions, by integrating out the CUTE modes $[\eta]$ with the
CUTE constants left over, $\{\sigma\}$,
\begin{eqnarray}
\label{CUTE1} Z\{[J_{EFT}]\}&\equiv& \int D[\phi]D[\eta] exp\{i
\int d^D x ({\mathcal{L}}_{CUTE}
([\phi([\eta])]|[\eta],[\sigma])+[J_{EFT} \ast
\phi([\eta])]\})\nonumber\\
&=& \int D[\phi_{\{\sigma\}}] exp\{i \int d^D x
({\mathcal{L}}_{EFT}([\phi_{\{\sigma\}}],[g]|\{\sigma\}) +[J_{EFT}
\ast \phi_{\{\sigma\}}])\},
\end{eqnarray}where $[J_{EFT}]$ denotes
the external sources for EFT and $[g]$ are EFT couplings (which
might be green functions in terms of CUTE parameters). In the
first line the path integral are naturally factorized as the
typical scales for EFT and CUTE should be widely separated.

The generating functional for vertex functions, upon field
expansion, now takes the following form,
\begin{eqnarray}
\label{GFEFT}
&&\Gamma_{EFT}([g],[\Phi_{\{\sigma\}}]|\{\sigma\})=\sum
\frac{1}{n!}\int \prod_{i=1}^n d^Dx_i
[\Phi_{\{\sigma\}}(x_i)]\Gamma^{(n)}([x_1,x_2,\cdots,x_n],[g]|\{\sigma\}),
\end{eqnarray}with
$\Gamma^{(n)}([x_1,x_2,\cdots,x_n],[g];\{\sigma\})$ being the
$n$-point complete vertex functions that are well defined in CUTE.
Removing the EFT loop contributions we get the EFT Lagrangian
(tree vertices),
\begin{eqnarray}
\label{tree} {\mathcal{L}}_{EFT}
([g],[\Phi_{\{\sigma\}}]|\{\sigma\})\equiv
\Gamma_{EFT}([g],[\Phi_{\{\sigma\}}]|\{\sigma\})|_{\text{loops
removed}}.
\end{eqnarray}
As no EFT loops are involved such EFT object is well defined in
the low energy (LE) limit ${\mathbf{L}}^{\{\sigma\}}$:
$\frac{\Lambda_{EFT}}{\Lambda_{CUTE}}\Longrightarrow 0$,
\begin{eqnarray}
\label{EFTTREE} {\mathcal{L}}_{EFT}([g],[\Phi])\equiv
{\mathbf{L}}^{\{\sigma\}}{\mathcal{L}}_{EFT}
([g],[\Phi_{\{\sigma\}}]|\{\sigma\}),
\end{eqnarray}that means the CUTE modes  or new physics are
completely decoupled from EFT at the Lagrangian or tree level (and
in certain convergent loop diagrams). But this does not
automatically lead to the complete decoupling in certain set of
EFT loops due to ill-definedness, because the low energy limit
operation and the EFT loop integration (summation over
intermediate states) do not commute on these amplitudes,
\begin{equation} [{\mathbf{L}}^{\{\sigma\}},\int_{\text{loop or
path}}]=[{\mathbf{L}}^{\{\sigma\}}, \sum_{\text{intermediate
states}}]\neq0.
\end{equation}The correct way is to carry out the loop integration
first\footnote{One might argue that the low energy limit might
again lead to UV divergences. To respond, we note that
ill-definedness is with our knowledge, but not with the nature.
Physical theories keep becoming better defined as more underlying
physics are uncovered, recall the evolution from Fermi's $\beta$
decay theory to the electroweak theory. In fact, there does exist
a mathematical framework that could yield finite results without
subtraction\cite{Epstein}. Here we wish to explore the physical
regularity masked by the ill-definedness that might be more
appealing to physicists.}.

One might contend that the preceding simple arguments seems
useless in practice as CUTE (or new physics) are unknown at all.
We disagree with this point of view. In next subsection we
demonstrate a further and, we think, nontrivial use of the
regularity of CUTE or its existence in computing the EFT loop
amplitudes.
\subsection{differential equations following from the existence of
underlying theory} Let us illustrate the method on a one loop
Feynman diagram $\gamma$ that is divergent in EFT parametrization.
In CUTE definition it should take the following form,
\begin{eqnarray}
\label{1loop} \Gamma_{\gamma} ([p],[g]|\{\sigma\})=\int d^4 k
f_{\gamma}(k,[p],[g]|\{\sigma\}).
\end{eqnarray} To determine its low energy limit we note the
following fact\cite{Diffr,Caswell}: Differentiating a divergent
amplitude with respect to external parameters (momenta and masses)
reduces and even removes the divergence. In CUTE language, an EFT
loop's dependence upon $\{\sigma\}$ is reduced or completely
removed (CUTE effects completely decoupled) if appropriate times
of differentiation are done. For the one loop example
$\Gamma_{\gamma} ([p],[g]|\{\sigma\})$, this is,
\begin{eqnarray}
&&\partial^{\omega_{\gamma}}_{[p]}\{{\mathbf{L}}^{\{\sigma\}}\Gamma_{\gamma}
([p],[g]|\{\sigma\})\}={\mathbf{L}}^{\{\sigma\}}\{ \int d^4 k
\partial^{\omega_{\gamma}}_{[p]}f_{\gamma}(k,[p],[g]|\{\sigma\})\}\nonumber \\
&=& \int d^4 k
\partial^{\omega_{\gamma}}_{[p]}\{{\mathbf{L}}^{\{\sigma\}}
f_{\gamma}(k,[p],[g]|\{\sigma\})\}= \int d^4 k
\partial^{\omega_{\gamma}}_{[p]} f_{\gamma}(k,[p],[g])\equiv
\Gamma^{(\omega_{\gamma})}_{\gamma}([p],[g]).
\end{eqnarray} The above derivation is based on and guaranteed by the
regularity of the CUTE definition, a nontrivial use of the
existence of CUTE as stated above. $\omega_{\gamma}-1$ is set
equal to the superficial degree of $\gamma$ according to Weinberg
theorem\cite{Weinb}. Then the loop integration can be safely
performed within EFT.

Technically, this method is to obtain the CUTE version of EFT
amplitudes from the following kind of inhomogeneous differential
equations,
\begin{eqnarray}
\label{DIFFEQ}
&&\partial^{\omega_{\gamma}}_{[p]}Y_{\gamma}([p],[g];\{C\})=
\Gamma^{(\omega_{\gamma})}_{\gamma}([p],[g]),
\end{eqnarray}with $\{C\}$ representing
the ambiguities inherent in differential equation solutions. So we
can only determine the quantities we want up to a polynomial,
\begin{eqnarray} &&{\mathbf{L}}^{\{\sigma\}}\Gamma_{\gamma}
([p],[g]|\{\sigma\}) =Y_{\gamma}([p],[g];\{C\}) \text{mod}
{\mathcal{P}}^{(\omega_{\gamma}-1)},
\end{eqnarray}with ${\mathcal{P}}^{(\omega_{\gamma}-1)}$ being the
polynomial. Of course the constants $\{C\}$ should be uniquely
defined in CUTE. But the validity of differential equations
derived above implies that: the decoupling effects from CUTE or
new physics must lie in the space spanned by $\{C\}$, {\em no
matter what they are}. In this sense, the differential equation
approach is the most general way for parametrizing the possible
form of CUTE or new physics in the decoupling limit.

In principle, fixing the constants $\{C\}$ according to physical
(boundary) conditions or experimental data, just like what we do
in classical electrodynamics and quantum mechanics\footnote{It is
easy to see that the constants thus determined must be, in the
conventional renormalization terminology, scheme and scale
invariant\cite{RS}.}, can spare the procedure of intermediate
subtractions and lead to a simple and natural strategy for
renormalization\cite{CUTE}. The application to multi-loop cases is
straightforward, see Ref.\cite{CUTE} for detailed description. A
nontrivial two loop example can be found in Ref.\cite{D8164}. The
annoying overlapping divergences are automatically dissolved in
this differential equation approach\cite{Caswell}. From now on we
temporarily call the strategy the differential equation approach.

Graphically, the differentiation is to insert scalar
($\partial_{[\text{mass}]}$) or vector ($
\partial_{[\text{momentum}]}$)  EFT vertex (with zero momentum
transfer $q=0$) into the diagrams, which lowers their degree of
superficial divergence, see Fig.1.
\begin{center}
\begin{picture}(310,120)(-160,-60)
\GOval(-50,0)(25,20)(0){1} \Vertex(-70,0){2} \Vertex(-30,0){2}
\LongArrow(-10,0)(10,0) \GOval(50,0)(25,20)(0){1} \Vertex(30,0){2}
\Vertex(70,0){2} \Vertex(50,25){3} \DashLine(50,40)(50,25){2}
\Text(-50,35)[]{p+k} \Text(-50,-30)[]{k}
\Text(20,35)[]{p+k}\Text(50,-30)[]{k} \Text(85,35)[]{p+k}
 \Text(50,50)[]{$q=0$}
\Text(-95, 0)[]{$\partial_p $} \Text(-10,-70)[b]{Fig.1.
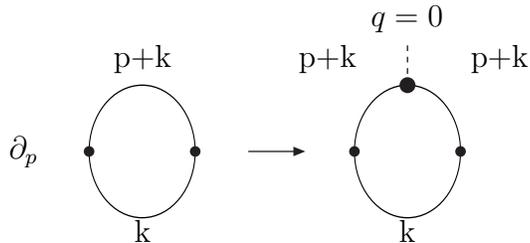
Differentiation w.r.t. external momentum}
\end{picture}
\end{center}
\subsection{Two simple examples}
First let us consider the following simple one  fermion loop
amplitude,
\begin{eqnarray}
\label{example} \Gamma_{1\nu}&=&2m\int \frac{d^4l}{(2\pi)^4}
{\mathrm{tr}}(\frac{i}{\gamma_{\alpha}l^{\alpha}-m}
\gamma_{\nu}\gamma_5\frac{i}{\gamma_{\alpha}
(l-p)^{\alpha}-m}\gamma_5)\nonumber \\
& =&-8m^2p_{\nu}\int \frac{d^4l}{(2\pi)^4}
\frac{1}{[l^2-m^2][(l-p)^2-m^2]}\equiv-8m^2p_{\nu}I.
\end{eqnarray}This loop has a linear superficial, twice
differentiation with respect to $p$ or $m$ are enough.

Technically, the differentiation can be performed either at the
graph level or at the reduced integral level (i.e., on $I$),
provided Lorentz invariance are required. Usually (see, Sec. III),
a fermion loop amplitude would give rise to a sum of integrals
with different divergences, some are even convergent, thus it will
be time consuming to perform the differentiation at the graph
level. In this example, the true divergence in $I$ is only
logarithmic, and once differentiation is OK. Therefore we choose
to differentiate at the level of the reduced integral ($I$),
\begin{eqnarray}
\partial_{p^{\mu}}I=2\int \frac{d^4l}{(2\pi)^4}
\frac{(l-p)_\mu}{[l^2-m^2][(l-p)^2-m^2]^2}\equiv 2I_{1;\mu}-2p_\mu
I_2.
\end{eqnarray}Now the two integrals $I_{1;\mu}$ and $I_2$ are
well defined within EFT and the loop integration can be done,
\begin{eqnarray}
I_{1;\mu}=-\frac{ip_\mu}{(4\pi)^2}\int^1_0 dx
\frac{(1-x)^2}{m^2+p^2(x^2-x)},\ \ \
I_2=-\frac{i}{(4\pi)^2}\int^1_0 dx \frac{1-x}{m^2+p^2(x^2-x)}.
\end{eqnarray}Then we have,
\begin{eqnarray}
\partial_{p^{\mu}}I=-2\frac{ip_\mu}{(4\pi)^2}\int^1_0 dx
\frac{(x^2-x)}{m^2+p^2(x^2-x)}=\partial_{p^{\mu}}
\{-\frac{i}{(4\pi)^2}\int^1_0 dx \ln (m^2+p^2(x^2-x)) \},
\end{eqnarray}from which the solution can be readily found as
\begin{eqnarray}
&&I=-\frac{i}{(4\pi)^2}[\int^1_0 dx \ln
\frac{m^2+p^2(x^2-x)}{\mu^2}+C],\\
\Longrightarrow
&&\Gamma_{1\nu}=\frac{im^2p_{\nu}}{2\pi^2}[\int^1_0 dx \ln
\frac{m^2+p^2(x^2-x)}{\mu^2}+C],
\end{eqnarray}with two unknown constants: one dimensional,$\mu$, the
other dimensionless $C$. It is easy to see that they should be
independent of the tree parameter $m$: $\partial_m \mu=\partial_m
C=0$.

As mentioned above, differentiation at the graph level should lead
to the same expression, provided Lorentz invariance is imposed on
the amplitude. Therefore, Lorentz non-invariant constants like
$C^\nu$ are discarded. Of course, under certain circumstances,
such kind of constants might be allowed and useful.

As the second example we calculate the following one loop
amplitude,
\begin{eqnarray}
&&\Gamma_2=4m^2\int\frac{d^4l}{(2\pi)^4}
{\mathrm{tr}}(\frac{i}{\gamma_{\alpha}l^{\alpha}-m}m
\frac{i}{\gamma_{\alpha}l^{\alpha}-m}\gamma_5
\frac{i}{\gamma_{\alpha}(l-p)^{\alpha}-m} \gamma_5)
=16im^4(I_3-p^2I_4),\nonumber \\ &&I_3
=\int\frac{d^4l}{(2\pi)^4}\frac{1}{(l^2-m^2)^2},\ \ \  I_4=\int
\frac{d^4l}{(2\pi)^4} \frac{1}{(l^2-m^2)^2((l-p)^2-m^2)}.
\end{eqnarray}Of the two integrals only $I_3$ has a logarithmic
divergence, $I_4$ is convergent. So we only need to differentiate
$I_3$ once, with respect to mass. However, noting that
$I_3=I|_{p=0}$, we could make use of the preceding calculation to
obtained that $I_3=-\frac{i}{(4\pi)^2}[ \ln \frac{m^2}{\mu^2}+C]
$. Then with $I_4=-\frac{i}{2(4\pi)^2}\int^1_0 dx
\frac{1}{m^2+p^2(x^2-x)}$ we have,
\begin{eqnarray}
\Gamma_2=\frac{m^2}{2\pi^2}
\{2m^2(\ln\frac{m^2}{\mu^2}+C)-{p^2}\int^1_0 dx
\frac{1}{m^2+p^2(x^2-x)}\}.
\end{eqnarray}Again, we stress that this is the most general
parametrization of the two amplitudes that are Lorentz invariant.
From the above examples we could also see that there is no
subtlety like the definition of metric tensor and $\gamma_5$ that
is associated with dimensional schemes\cite{Jeg}.
\section{Calculation and Parametrization of two point functions of axial
current and its divergence} For simplicity we work in $U(1)$ gauge
theory with just one massive fermion, i.e., QED. The objects to be
computed and investigated are respectively\cite{PCDC}
\begin{eqnarray}
\label{def1} &&{\Pi}^5_{\mu\nu}(p,-p)\equiv i
{\mathcal{FT}}\{\langle 0|T(j_{5\mu}j_{5\nu})|0\rangle\},\ \
\Delta^5_{\mu\nu}(0,p,-p)\equiv{\mathcal{FT}}\{\langle 0|T(\theta
j_{5\mu}j_{5\nu})|0\rangle\};\\
 &&{\Pi}^5_{\ \nu}(p,-p)\equiv i{\mathcal{FT}}\{\langle
0|T(j_{5}j_{5\nu})|0\rangle\},\ \ \Delta^5_{\
\nu}(0,p,-p)\equiv{\mathcal{FT}}\{\langle 0|T(\theta
j_{5}j_{5\nu})|0\rangle\}; \\
&&{\Pi}^5(p,-p)\equiv i{\mathcal{FT}}\{\langle
0|T(j_{5}j_{5})|0\rangle\},\ \
\Delta^5(0,p,-p)\equiv{\mathcal{FT}}\{\langle 0|T(\theta
j_{5}j_{5})|0\rangle\};\\
&&\langle\sigma\rangle\equiv4m\langle\bar{\psi}\psi\rangle,\ \
\Pi^{\theta\sigma}(0,0)\equiv
-i{\mathcal{FT}}\{\langle0|\theta\sigma|0\rangle\},\\
&&j^5_{\mu}\equiv \bar{\psi}\gamma_{\mu}\gamma_5\psi,\ \
j^5\equiv2im\bar{\psi}\gamma_5\psi, \ \ \theta\equiv
m\bar{\psi}\psi,\ \ \ \sigma\equiv 4m\bar{\psi}\psi.
\end{eqnarray}where ${\mathcal{FT}}\{\cdots\}$ denotes the Fourier
transform and $m$ refers to the fermion mass. At one loop level
they read (see Fig.1)
\begin{eqnarray}
\label{def2} &&{\Pi}^5_{\mu\nu}(p,-p)=-i\int \frac{d^4l}{(2\pi)^4}
{\mathrm{tr}}(\frac{i}{A}\gamma_{\nu}\gamma_5\frac{i}{B}
\gamma_{\mu}\gamma_5),\\
&&\Delta^5_{\mu\nu}(0,p,-p)= -\int \frac{d^4l}{(2\pi)^4}
{\mathrm{tr}}(\frac{i}{A}m
\frac{i}{A}\gamma_{\nu}\gamma_5\frac{i}{B}
\gamma_{\mu}\gamma_5) +\text{cross term};\\
 &&{\Pi}^5_{\ \nu}(p,-p)=2m\int \frac{d^4l}{(2\pi)^4}
{\mathrm{tr}}(\frac{i}{A}\gamma_{\nu}\gamma_5\frac{i}{B}\gamma_5), \\
&&\Delta^5_{\ \nu}(0,p,-p)=-2im\int \frac{d^4l}{(2\pi)^4}
{\mathrm{tr}}(\frac{i}{A}m
\frac{i}{A}\gamma_{\nu}\gamma_5\frac{i}{B}
\gamma_5) +\text{cross term}; \\
&&{\Pi}^5(p,-p)=4m^2i\int \frac{d^4l}{(2\pi)^4}
{\mathrm{tr}}(\frac{i}{A}\gamma_5\frac{i}{B}\gamma_5),\\
&&\Delta^5(0,p,-p)=4m^2\int\frac{d^4l}{(2\pi)^4}
{\mathrm{tr}}(\frac{i}{A}m \frac{i}{A}\gamma_5\frac{i}{B}
\gamma_5) +\text{cross term} ;\\
&&\langle\sigma\rangle=-4m \int
\frac{d^4l}{(2\pi)^4}{\mathrm{tr}}(\frac{i}{A}),\ \
\Pi^{\theta\sigma}(0,0)= 4im^2\int
\frac{d^4l}{(2\pi)^4}{\mathrm{tr}}(\frac{i}{A}\frac{i}{A}),\\
&&A\equiv \gamma_{\alpha}l^{\alpha}-m,\ \ B\equiv
\gamma_{\alpha}(l-p)^{\alpha}-m.
\end{eqnarray}
It is easy to carry out the Dirac traces to arrive at the
following form
\begin{eqnarray}
\label{def3} {\Pi}^5_{\mu\nu}(p,-p)&=&2i
g_{\mu\nu}((p^2-4m^2)I_{ab}-I_a-I_b) +8iI_{ab;\mu\nu}-4i
p_{\mu}I_{ab;\nu} -4ip_{\nu}I_{ab;\mu}
,\\
\Delta^5_{\mu\nu}(0,p,-p)&=&8im^2 g_{\mu\nu}(p^2I_{a^2b}-I_{a^2}-
2 I_{ab}-4m^2I_{a^2b})+32im^2I_{a^2b;\mu\nu}\nonumber \\
&&-16im^2(p_{\mu}I_{a^2b;\nu}+p_{\nu}I_{a^2b;\mu});\\
{\Pi}^5_{\ \nu}(p,-p)&=-&8m^2I_{ab}p_{\nu},\\
\Delta^5_{\ \nu}(0,p,-p)&=&16m^2I_{ab;\nu}-
16m^2(I_{ab}+2m^2I_{a^2b})p_{\nu}; \\
{\Pi}^5(p,-p)&=&8im^2(I_a+I_b-p^2I_{ab}),\\
\Delta^5(0,p,-p)&=&32im^4(I_{a^2}-p^2I_{a^2b}); \\
\langle\sigma\rangle&=&-16im^2I_a,\ \ \ \Pi^{\theta\sigma}(0,0)=
\langle\sigma\rangle-32im^4I_{a^2},\\
 I_{a^nb^k}&\equiv&\int \frac{d^4l}{(2\pi)^4} \frac{1}{a^nb^k},\ \
I_{a^nb^k;\mu}\equiv\int \frac{d^4l}{(2\pi)^4}
\frac{l_{\mu}}{a^nb^k},I_{a^nb^k;\mu\nu}\equiv\int
\frac{d^4l}{(2\pi)^4}
\frac{l_{\mu}l_{\nu}}{a^nb^k},\\
 a&\equiv& l^2-m^2,\ \ b\equiv (l-p)^2-m^2.
\end{eqnarray}
Most of the integrals are UV ill-defined and hence can not be
directly performed.

To calculate these integrals we will employ two different methods
as mentioned in the introduction, one is the most frequently used
dimensional regularization, another is the differential equation
approach introduced in sec. II. It is easy to perform the
calculations in the two approaches and the parametrization of the
integrals will be given in the appendix A. With these integrals
the two- and three- point functions read
\begin{eqnarray}
{\Pi}^5_{\mu\nu}(p,-p)&=&\frac{2g_{\mu\nu}}{(4\pi)^2} \{
p^2[2C_0-C_4+\Delta_0+2\int^1_0 da (x^2-x)(\ln
\frac{D}{\mu^2}+C_4-1)]\nonumber \\
&&-2m^2[\Delta_0+\ln\frac{m^2}{\mu^2}+2C_0]+2C^{\prime}+2C_5^{\prime}\}
\nonumber \\
&&+\frac{p_{\mu}p_{\nu}}{4\pi^2}\{2\int^1_0 da (1-x)^2[\ln
 \frac{D}{\mu^2}+C_4]-C_0-\Delta_0\},\\
\Delta^5_{\mu\nu}(0,p,-p)&=&\frac{m^2p_{\mu}p_{\nu}}{\pi^2p^2}
(1-\frac{m^2}{\Delta})- \frac{g_{\mu\nu}m^2}{2\pi^2}
\{\ln\frac{m^2}{\mu^2}+2C_0+\Delta_0 +\frac{2m^2-p^2/2}{\Delta}
\};\\
{\Pi}^5_{\
\nu}(p,-p)&=&\frac{im^2p_{\nu}}{2\pi^2}(\Delta_0+C_0),\\
\Delta^5_{\ \nu}(0,p,-p)
&=&\frac{im^2p_{\nu}}{2\pi^2}(\Delta_0+C_0+\frac{2m^2}{\Delta});\\
{\Pi}^5(p,-p)&=&\frac{m^2}{2\pi^2}
\{2m^2(\ln\frac{m^2}{\mu^2}+C_0-1)+2C^{\prime}-p^2(\Delta_0+C_0)
\},\\
\Delta^5(0,p,-p)&=&\frac{m^2}{\pi^2}
\{2m^2(\ln\frac{m^2}{\mu^2}+C_0)-\frac{p^2}{\Delta}\};\\
\langle\sigma\rangle&=&\frac{m^2}{\pi^2}
\{m^2(1-\ln\frac{m^2}{\mu^2}-C_0)-C^{\prime}\},
\\
\Pi^{\theta\sigma}(0,0)&=& \frac{m^2}{\pi^2}
\{3m^2(1/3-\ln\frac{m^2}{\mu^2}-C_0)- C^{\prime}\}.
\end{eqnarray}

For comparison we also list the results in dimensional
regularization here,
\begin{eqnarray}
{\Pi}^{5,\epsilon}_{\mu\nu}(p,-p)&=&\frac{2g_{\mu\nu}}{(4\pi)^2}
\{ p^2[\Delta_0-\Gamma(\epsilon)+2\int^1_0 da (x^2-x)(\ln
\frac{D}{4\pi\mu^2}-\Gamma(\epsilon)-1)]\nonumber \\
&&-2m^2[\Delta_0+\ln\frac{m^2}{4\pi\mu^2}-2\Gamma(\epsilon)]\}\nonumber \\
&&+\frac{p_{\mu}p_{\nu}}{4\pi^2}\{2\int^1_0 da (1-x)^2[\ln
 \frac{D}{4\pi\mu^2}-\Gamma(\epsilon)]+\Gamma(\epsilon)-\Delta_0\},\\
\Delta^{5,\epsilon}_{\mu\nu}(0,p,-p)&=&\frac{g_{\mu\nu}m^2}{2\pi^2}
\{2\Gamma(\epsilon)-\ln\frac{m^2}{4\pi\mu^2}-\Delta_0
+\frac{p^2/2-2m^2}{\Delta}
\}+\frac{m^2p_{\mu}p_{\nu}}{\pi^2p^2}(1-\frac{m^2}{\Delta});\\
{\Pi}^{5,\epsilon}_{\
\nu}(p,-p)&=&\frac{im^2p_{\nu}}{2\pi^2}(\Delta_0-\Gamma(\epsilon)),\\
\Delta^{5,\epsilon}_{\ \nu}(0,p,-p)
&=&\frac{im^2p_{\nu}}{2\pi^2}(\Delta_0-\Gamma(\epsilon)+\frac{2m^2}{\Delta});\\
{\Pi}^{5,\epsilon}(p,-p)&=&\frac{m^2}{2\pi^2}
\{2m^2(\ln\frac{m^2}{4\pi\mu^2}-\Gamma(\epsilon)-1)
-p^2(\Delta_0-\Gamma(\epsilon))
\},\\
\Delta^{5,\epsilon}(0,p,-p)&=&\frac{m^2}{\pi^2}
\{2m^2(\ln\frac{m^2}{4\pi\mu^2}-\Gamma(\epsilon))-
\frac{p^2}{\Delta}\};\\
\langle\sigma\rangle^{\epsilon}&=&\frac{m^4}{\pi^2}
(\Gamma(\epsilon)+1-\ln\frac{m^2}{4\pi\mu^2}),\\
\Pi^{\theta\sigma,\epsilon}(0,0)&=&\frac{3m^4}{\pi^2}
(\Gamma(\epsilon)+1/3-\ln\frac{m^2}{\mu^2}).
\end{eqnarray}It is easy to see that setting
$C_0=C_4=-\Gamma(\epsilon), C^{\prime}=C_5^{\prime}=0$ in the
differential equation solutions we will arrive at dimensional
regularization for these integrals. Again we see that the
differential equation approach is a general parametrization. Now
we turn to the next section to study the trace and chiral
identities.
\section{Trace and chiral Ward identities and their determinants}
\subsection{Canonical identities and general parametrization of anomalies}
First Let us write down the canonical identities for
trace relation and chiral symmetry that should satisfied by the
above vertex functions\cite{PCDC}:
\begin{eqnarray}
\label{traceid1}
&&\Delta^5_{\mu\nu}(0,p,-p)=(2-p\partial_p){\Pi}^5_{\mu\nu}(p,-p),\\
\label{traceid2} &&\Delta^5_{\
\nu}(0,p,-p)=(2-p\partial_p){\Pi}^5_{\ \nu}(p,-p),\\
\label{traceid3}&&\Delta^5(0,p,-p)=(2-p\partial_p){\Pi}^5(p,-p);\\
\label{chiralid1} &&-ip^{\mu}\Delta^5_{\mu\nu}(0,p,-p)=\Delta^5_{\
\nu}(0,p,-p)+{\Pi}^5_{\ \nu}(p,-p),\\
\label{chiralid2}&&ip^{\nu}\Delta^5_{\
\nu}(0,p,-p)=\Delta^5(0,p,-p) +{\Pi}^5(p,-p)+
\Pi^{\theta\sigma}(0,0),\\
\label{chiraldi3}&&-ip^{\mu}{\Pi}^5_{\mu\nu}(p,-p)={\Pi}^5_{\
\nu}(p,-p),\\
\label{chiraldi4}&&ip^{\nu}{\Pi}^5_{\ \nu}(p,-p)={\Pi}^5(p,-p)
+\langle\sigma\rangle.
\end{eqnarray}The first three are trace identities and the other
four are chiral Ward identities.

Now let us check them with our results given above. Again we
should stress that in the differential equation approach, we have
parametrize the regularization dependence or CUTE decoupling
effects in a most general way that is consistent with Lorentz
invariance. If one choose not to obey Lorentz invariance, then
more constants that not Lorentz invariant like $C_{\mu},
C_{\mu\nu}$ should appear. This might be possible choices under
certain circumstances. Here we assume there is no violation of
Lorentz invariance in any form of CUTE and its low energy limits.
After simple algebra, we find that in the general parametrization
of the CUTE decoupling (or low energy limit) effects,
Eq.s(~\ref{traceid1},~\ref{traceid2},~\ref{traceid3},~\ref{chiraldi3})
are modified as follows,
\begin{eqnarray}
\label{atraceid1}
&&\Delta^5_{\mu\nu}(0,p,-p)=(2-p\partial_p){\Pi}^5_{\mu\nu}(p,-p)
+\frac{1}{6\pi^2}(g_{\mu\nu}p^2-p_{\mu}p_{\nu})-
\frac{g_{\mu\nu}}{\pi^2}(m^2+(C^{\prime}+C_5^{\prime})/2),\\
\label{atraceid2} &&\Delta^5_{\
\nu}(0,p,-p)=(2-p\partial_p){\Pi}^5_{\ \nu}(p,-p)+
\frac{im^2p_{\nu}}{\pi^2},\\
\label{atraceid3}&&\Delta^5(0,p,-p)=(2-p\partial_p){\Pi}^5(p,-p)
-\frac{m^2p^2}{\pi^2}+\frac{2m^2}{\pi^2}(m^2-C^{\prime});\\
\label{achiraldi3}&&-ip^{\mu}{\Pi}^5_{\mu\nu}(p,-p)={\Pi}^5_{\
\nu}(p,-p)-\frac{i(C^{\prime}+C_5^{\prime})p_{\nu}}{4\pi^2}.
\end{eqnarray}

Here we see that anomalies appear in all the three trace
identities and one chiral Ward identities. But chiral anomaly in
Eq.(~\ref{achiraldi3}) is regularization scheme or underlying
effects dependent, as long as $C^{\prime}+C_5^{\prime}=0$ the
chiral symmetry is restored in the two point functions, that is,
not all regularization schemes violate Eq.(~\ref{chiraldi3}), or
equivalently, such anomaly is sensitively dependent on the
underlying theory details, and the existence of such anomaly is
consistent with the trace identities with anomalies, see below.
This is a modification of Wilson's argument for two point
functions\cite{PCDC}. However, the anomaly in the trace identities
can not be removed in any regularization, or equivalently, the
CUTE decoupling effects do violate the EFT trace identities.
\subsection{Chiral anomaly and consistency}
Now let us check the consistency among the chiral and trace
identities with anomalies. First let us consider the case where
there is no chiral anomaly in Eq.(~\ref{chiraldi3}), i.e.,
$C^{\prime}+C_5^{\prime}=0$. Then applying the relations
Eq.s(~\ref{chiralid1},~\ref{chiralid2},~\ref{chiraldi3},~\ref{chiraldi4})
to the trace identities (~\ref{atraceid1},~\ref{atraceid2}), we
get
\begin{eqnarray}
\label{atraceid10}
&&\Delta^5_{\mu\nu}(0,p,-p)=(2-p\partial_p){\Pi}^5_{\mu\nu}(p,-p)
+\frac{1}{6\pi^2}(g_{\mu\nu}p^2-p_{\mu}p_{\nu})-
\frac{g_{\mu\nu}m^2}{\pi^2},\\
\label{atraceid20} &&\Delta^5_{\
\nu}(0,p,-p)=(2-p\partial_p){\Pi}^5_{\ \nu}(p,-p)+
\frac{im^2p_{\nu}}{\pi^2},\\
\label{atraceid30}&&\Delta^5(0,p,-p)=(2-p\partial_p){\Pi}^5(p,-p)
-\frac{m^2p^2}{\pi^2}+3\langle\sigma\rangle-\Pi^{\theta\sigma}(0,0).
\end{eqnarray} The second trace relation agrees exactly with
Ref.\cite{PCDC}, there are some disagreements in the other two:
(1) in Ref.\cite{PCDC}, the numerical coefficient of the anomaly
term $(g_{\mu\nu}p^2-p_{\mu}p_{\nu})$ is $\frac{1}{8\pi^2}$ while
in our first equation it is $\frac{1}{6\pi^2}$; (2) in
Ref.\cite{PCDC} the last two terms in Eq.(~\ref{atraceid30}) were
missing, we will discuss about this point later. Note that in
Eq.(~\ref{atraceid30}), $\frac{2m^2}{\pi^2}(m^2-C^{\prime})=
3\langle\sigma\rangle-\Pi^{\theta\sigma}(0,0)$ is used. We should
remind that this term can not be removed by setting
$C^{\prime}=m^2$ as the constant $C^{\prime}$ is independent of
$m$: $\partial_mC^{\prime}=0$. Thus putting
$C^{\prime}+C_5^{\prime}=0$ we can conclude that in any CUTE whose
decoupling limit preserve the chiral Ward identities for two point
functions, the trace anomalies at one loop level for the two point
functions are consistently given in
Eq.s(~\ref{atraceid10},~\ref{atraceid20},~\ref{atraceid30}), no
matter what numerical values $C_0, C_4$ may take. Or one can claim
that as long as the new physics in the low energy limit  does not
violate chiral symmetry in the two point functions, then the trace
relation at one loop level for these two point functions are
consistently dictated by these three equations.

Now let us consider the case where $C^{\prime}+C_5^{\prime}\neq0$.
Imposing the chiral relations
Eq.s(~\ref{chiralid1},~\ref{achiraldi3}) on Eq.(~\ref{atraceid1})
we have,
\begin{eqnarray}
\label{wrong1} &&\Delta^5_{\
\nu}(0,p,-p)=(2-p\partial_p){\Pi}^5_{\ \nu}(p,-p)+
\frac{im^2p_{\nu}}{\pi^2},
\end{eqnarray}which is exactly Eq.(~\ref{atraceid2}). Then imposing
Eq.s(~\ref{chiralid2},~\ref{chiraldi4}) on Eq.(~\ref{atraceid2})
we find that
\begin{eqnarray}
\label{wrong2}&&\Delta^5(0,p,-p)=(2-p\partial_p){\Pi}^5(p,-p)
-\frac{m^2p^2}{\pi^2}+
3\langle\sigma\rangle-\Pi^{\theta\sigma}(0,0),
\end{eqnarray}again in agreement with Eq.(~\ref{atraceid3}) or Eq.
(~\ref{atraceid30}). This completes the consistency check on the
identities (Eq.s (~\ref{atraceid1}), (~\ref{atraceid2}),
(~\ref{atraceid3}), (~\ref{chiralid1}), (~\ref{chiralid2}),
(~\ref{achiraldi3}) and (~\ref{chiraldi4})) in the most general
parametrization of the decoupling effects from CUTE or new
physics.

In other words, in the regularization schemes or CUTE decoupling
limit where $C^{\prime}+C_5^{\prime}\neq0$, the chiral Ward
identities (with anomalies) are also consistent with the trace
identities (with anomalies) for the two point functions, at least
at one loop level. It is not necessary to impose chiral symmetry
in the Ward identities (~\ref{chiralid1}), (~\ref{chiralid2}),
(~\ref{chiraldi3}) and (~\ref{chiraldi4}) so that
$C^{\prime}+C_5^{\prime}=0$. That is the chiral schemes (where
$C^{\prime}+C_5^{\prime}=0$) are not mathematically superior to
the ones violating chiral symmetry (where
$C^{\prime}+C_5^{\prime}\neq0$). Of course, the final
determination of what value $C^{\prime}+C_5^{\prime}$ should take
must be searched in physical 'data'.

Obviously dimensional regularization is a scheme preserving chiral
symmetry, where $C^{\prime}+C_5^{\prime}=0$. In fact
$C^{\prime}=C_5^{\prime}=0$ in dimensional regularization. It is
just a special way of parametrizing the divergence which might
become unfavorable and should be altered in certain
situations\cite{KSW}, for the correspondence between the DR poles
and the cutoff powers, see Ref.\cite{Veltman}.
\subsection{Remarks on Eq.(~\ref{atraceid30})}
Now let us return to Eq.(~\ref{atraceid30}) or (~\ref{atraceid3}).
From the trace identity perspective, both $-\frac{m^2p^2}{\pi^2}$
and $3\langle\sigma\rangle-\Pi^{\theta\sigma}(0,0)$ are anomalies
(we have explained in the preceding subsection that the latter
could not be removed by setting $C^{\prime}=m^2$). But the latter
is required by and explicable within chiral Ward identities and
its existence is independent of CUTE or new physics, thus we
arrive at an interesting phenomenon: {\em the canonical terms in
chiral identity become anomalies in trace identity}. Thus it seems
that EFT chiral symmetry is quite different with EFT scaling laws:
the former could be preserved by new physics or CUTE (i.e., new
physics or CUTE heavy modes could be completely decoupled from EFT
chiral identities ($C^{\prime}+C_5^{\prime}=0$) provided the
underlying contents are free of chiral anomaly
everywhere\cite{Hooft}), but the latter is inevitably modified by
CUTE or new physics, that is, CUTE or new physics can never be
completely decoupled from the EFT scaling laws even when chiral
symmetry is perfectly preserved. This might lead to the
speculation that CUTE or new physics and EFT's might not differ
too much in the chiral symmetry perspective, but they should be
essentially different in the scale symmetry.

As our investigation is done in a non-supersymmetric EFT, it would
be interesting to see how this disparity between chiral and scale
symmetry evolves as the theory turns supersymmetric, where the
dilatation current and axial vector current consist a
supersymmetric multiplet and should have the identical anomaly
status\cite{SV}.

This disparity might also be understood somehow from the
non-renormalization of chiral anomaly which is known as the
Adler-Bardeen theorem\cite{AB}. In CUTE language, this means that
the CUTE modes only slightly modify the EFT chiral symmetry at the
one loop level, they are completely decoupled from chiral symmetry
in all the rest EFT amplitudes, or they are even completely
decoupled in appropriate EFT contents\cite{Hooft}. But the CUTE
modes significantly modify the EFT trace identities or scaling
laws in all orders of the EFT loop amplitudes (for certain
vertices) that are ill-defined, no matter to what degree chiral
symmetry is affected. Thus the non-renormalization theorems of
various objects and relations might lead to similar phenomenon,
more examples might be found in supersymmetric
contexts\cite{weinsusy}.
\section{Interpretation of the results}
To interpret the trace anomalies in CUTE language, we first note
that the full scaling law (trace identities) in CUTE version of
the EFT vertex should read,
\begin{eqnarray}
\Gamma^{(n)}([\lambda p],[\lambda^{d_g} g]; \{
\lambda^{d_{\sigma}} \sigma\})= \lambda^{d_{\Gamma^{(n)}}}
\Gamma^{(n)} ([p],[g];\{\sigma\}),
\end{eqnarray}
where $d_{\cdots}$ refers to the canonical scale dimension of a
parameter or canonical mass dimension of a constant (effective
and/or underlying), that is the vertex function {\bf must} be a
homogeneous function of all its dimensional arguments. In
differential form, it is
\begin{eqnarray}
\label{CUTEscaling} \{ \sum_{i=1}^n p_i \cdot \partial_{p_i} +
\sum d_{\sigma} {\sigma}
\partial_{\sigma} +\sum d_g g \partial_g -d_{\Gamma^{(n)}} \}
\Gamma^{(n)}([p],[g];\{\sigma\})=0,
\end{eqnarray}where $d_{\cdots}$ refers to the canonical scaling
dimension.

Specifically, for the one-loop two point functions considered
above Eq.(~\ref{CUTEscaling}) implies that
\begin{eqnarray} \label{QEDscaling} &&\{ p \partial_{p}
+ \sum d_{\sigma} {\sigma}
\partial_{\sigma} +m \partial_m -2 \}
\Pi^{\cdots}(p, -p,m;\{\sigma\})=0\nonumber \\
\Longrightarrow &&  \Delta^{\cdots} (p,0,-p, m;\{\sigma\})=(2-p
\partial_{p} - \sum d_{\sigma} {\sigma}\partial_{\sigma})
\Pi^{\cdots}(p, -p,m;\{\sigma\}),
\end{eqnarray}where $\Delta^{\cdots} (p,0,-p, m;\{\sigma\})\equiv m
\partial_m
\Pi^{\cdots}(p, -p,m;\{\sigma\})$. Here, we have explicitly put
mass $m $ into the arguments of the two- and three- point
functions. Obviously, Eq.s (~\ref{CUTEscaling},~\ref{QEDscaling})
describe the 'canonical' scaling law (or trace identities) in CUTE
language, no anomalies. Then, in the decoupling limit, if $\sum
d_{\sigma} {\sigma}\partial_{\sigma}$ 's contributions vanish, we
say a complete decoupling of CUTE modes is realized and there is
no anomaly in the EFT identities. However, as is  shown above, at
least for the two point functions under consideration, this term
leads to a non-vanishing polynomial in terms of the EFT parameters
in the decoupling limit, or the local operators in terms of EFT
fields, which, unexpected from the EFT deduction, become anomalies
for EFT language:
\begin{eqnarray} &&{\mathbf{L}}^{\{\sigma\}}\{\sum d_{\sigma}
{\sigma}\partial_{\sigma}\Pi^{\cdots}(p, -p,m;\{\sigma\})\}= \sum
d_{\bar{c}} {\bar{c}}\partial_{\bar{c}}\Pi^{\cdots}(p,
-p,m;\{\bar{c}\})=\text{anomalies},\\
&&\Pi^{\cdots}(p, -p,m;\{\bar{c}\})={\mathbf{L}}^{\{\sigma\}}
\Pi^{\cdots}(p, -p,m;\{\sigma\}),
\end{eqnarray}where $\{\bar{c}\}$ refers to the constants defined
by the decoupling limit, a CUTE definition of the constants
$\{C\}$ that are obtained from differential equations in EFT.

Noting that the non-decoupling of CUTE modes just implies the UV
divergence or ill-definedness in EFT, we technically reproduced
the conventional interpretation that trace anomalies or violation
of scaling laws are due to UV divergence. As UV divergence is
unphysical, here we propose a physical rationale for the anomaly
phenomenon by providing a general parametrization of the CUTE or
new physics decoupling effects, which accommodates all possible
regularization schemes, as long as they are mathematically
consistent. Thus, the dimensional transmutation is nothing but the
transformation of the 'canonical' CUTE structures or new physics
into the EFT anomalies in the decoupling limit.

Considering the fact that all the decoupling effects in a loop
amplitude always appear in the local (polynomial in momentum
space) part of each loop amplitude of a vertex, we can easily
arrive at the following operatoral equality: \begin{eqnarray}
\label{operatorRGE} &&
{\mathbf{L}}^{\{\sigma\}}\{\sum_{\{\sigma\}}
d_{\sigma}\sigma\partial_{\sigma}\}= \sum_{\{\bar c\}} d_{\bar c}
{\bar c} \partial_{\bar c}=\sum_{\{{\mathcal{O}}_{i}\}}
{\delta}_{{\mathcal{O}}_{i}}\hat{I}_{{\mathcal{O}}_i} =\text{trace
or scale anomalies}\end{eqnarray} as $\sum_{\{\bar c\}} d_{\bar
c}{\bar c}
\partial_{\bar c}$ always induce the insertion of the local
operators ($\{{\mathcal{O}}_i\}$) corresponding to the vertex
functions in each loop component of any EFT Feynman diagram. Here
it is easy to see that ${\delta}_{{\mathcal{O}}_{i}}$ should be
the 'anomalous' dimension of the composite operator
${{\mathcal{O}}_{i}}$. Noting that parametrizing the constants
$\{\bar{c}\}$ in terms of an independent scale $\mu$ and a number
of dimensionless constants $\{\bar {c_0}\}$,
Eq.(~\ref{operatorRGE}) is in fact an operator form for RGE in
CUTE language.
\begin{eqnarray}
\label{operatorRGE1} {\mathbf{L}}^{\{\sigma\}}\{\sum_{\{\sigma\}}
d_{\sigma}\sigma\partial_{\sigma}\}= \mu
\partial_{\mu}=\sum_{\{{\mathcal{O}}_{i}\}}
{\delta}_{{\mathcal{O}}_{i}}\hat{I}_{{\mathcal{O}}_{i}}
\Longrightarrow  \mu
\partial_{\mu}-\sum_{\{{\mathcal{O}}_{i}\}}
{\delta}_{{\mathcal{O}}_{i}}\hat{I}_{{\mathcal{O}}_{i}}=0.\end{eqnarray}
Note that here all the tree parameters are 'bare' and 'physical'
in the sense that they should be defined by CUTE or physical
boundary conditions or data, unlike in the conventional
renormalization programs. Moreover, the composite operators
$\{{\mathcal{O}}_{i}\}$ might or might not come from EFT
Lagrangian, thus this equation is generally valid in any EFT,
whether it is renormalizable or not in the conventional
sense\cite{8111}.

For the EFT generating functional defined in CUTE
(Eq.(~\ref{GFEFT})), we can easily write down the following
scaling law or trace identity with the help of CUTE
\begin{eqnarray}
\label{1PIscaling} &&\{\sum_{\{ \Phi_{\{\sigma\}}\}}\int d^D x
[(d_{\Phi_{\{\sigma\}}}-x\cdot\partial_x) \Phi_{\{\sigma\}}(x)]
\frac {\delta}{\delta \Phi_{\{\sigma\}}(x)} + \sum d_gg\partial_g
+\sum d_{\sigma} {\sigma}\partial_{\sigma}-D\}\nonumber \\
&& \times \Gamma_{EFT}([g],[\Phi_{\{\sigma\}}]|\{\sigma\})=0,  \\
\label{1PIscaling2}
 & &\Longrightarrow\{\sum_{\{\Phi\}}\int d^D x
[(d_{\Phi}-x\cdot\partial_x) \Phi(x)] \frac {\delta}{\delta
\Phi(x)} + \sum_{[g]}d_gg\partial_g +\sum_{\{\bar c\}}
d_{\bar c} {\bar c} \partial_{\bar c}-D\}  \nonumber \\
& &\times \Gamma^{1PI} ([\Phi],[g];\{\bar c\}) =0.
\end{eqnarray}
with $D$ denoting the spacetime dimension. This equation holds for
any consistent EFT's. Since $\sum_{[g]}d_gg\partial_g
=\text{canonical trace tensor}=g_{\mu\nu}\Theta^{\mu\nu}$ in any
EFT, then in operator form, Eq.(~\ref{1PIscaling2}) reads
\begin{eqnarray}
\label{generaltrace}
&&g_{\mu\nu}\hat{{\Theta}}^{\mu\nu}=\text{canonical
trace}+\sum_{{\mathcal{O}}_{i}}
{\delta}_{{\mathcal{O}}_{i}}{\mathcal{O}}_{i},
\end{eqnarray}where $\sum_{\{{\mathcal{O}}_{i}\}}
{\delta}_{{\mathcal{O}}_{i}}{{\mathcal{O}}_{i}}$ consist the trace
anomaly coming from the decoupling limit of the 'canonical' CUTE
'trace' term: ${\mathbf{L}}^{\{\sigma\}}\sum_{\{\sigma\}}
d_{\sigma}\sigma \partial_{\sigma}$. Thus the formal derivation of
trace anomaly\cite{Trace} in the CUTE language is very simple. The
remaining work is to find the relevant loop diagrams for the
$n$-point functions that are ill-defined in EFT.

Similarly, we could anticipate that, for chiral identities in CUTE
version, we might have, for example, corresponding to
Eq.(~\ref{chiraldi3}),
\begin{eqnarray}
&&ip^{\mu}{\Pi}^5_{\mu\nu}(p,-p|\{\sigma\})={\Pi}^5_{\
\nu}(p,-p|\{\sigma\})+\text{terms normal in CUTE},
\end{eqnarray}which becomes, in the decoupling limit, the
following form,
\begin{eqnarray}
&&ip^{\mu}{\Pi}^5_{\mu\nu}(p,-p)={\Pi}^5_{\
\nu}(p,-p)+{\mathbf{L}}^{\{\sigma\}} \{\text{terms normal in
CUTE}\}.
\end{eqnarray} The last term, if does not vanish in the decoupling
limit, constitutes the anomaly to the chiral identity. This is
just the CUTE extension of the fact that decoupling a heavy
fermion leads to chiral anomaly in the triangle diagrams. In
regularization language, the Pauli-Villars regulator could serve
as a crude but intuitive substitute for the CUTE decoupling
interpretation of chiral anomaly\cite{PCAC}. In gauge theories
without chiral coupling and in $\lambda\phi^4$\cite{NTrace}, the
trace anomaly could also be obtained through heavy matter
decoupling\cite{Appel}.

Now we might speculate that all the EFT anomalies could be
understood from the perspective of the underlying theory or new
physics. Working with the differential equation approach, one
could have a general parametrization of all the possible
underlying theory's effects and discuss the issues without the
limitations associated with a specific regularization to arrive at
general conclusions.
\section{Discussions and summary}
Now let us address some important issues.

Although the philosophy followed here is well known, we make
further use of the philosophy in that we interpret it as the
existence of a complete formulation of the theory that underlies
all the EFT's and is well-defined in all respects, especially in
the high energy region that is above the EFT domains. From this
interpretation, one could derive a natural differential equation
approach for computing the loop diagrams, which is in fact a
general way for parametrizing all the possible underlying theory's
decoupling effects (or all possible consistent regularization
schemes).

In this sense, though each regularization is artificial and
defected somehow, the necessity in employing one in calculating
the divergent loops (or the necessity in introducing subtractions,
say in BPHZ, which is often viewed as regularization independent),
can be naturally understood in CUTE as a profound and physical
phenomenon: the non-decoupling of the CUTE details in the EFT
loops, though they are completely decoupled from the EFT
Lagrangian. The origin of UV divergence is identified in the CUTE
language (or the EFT philosophy) as the non-commutativity of the
operation of decoupling the CUTE modes and the EFT loop
integration (or equivalently the summation over EFT intermediate
states).

In such conception, we can easily discuss the anomalies (in the
two point functions under consideration) in a general
parametrization, and the anomalies can be naturally interpreted as
the non-vanishing decoupling effects of the 'canonical' CUTE
contributions. Moreover, we have shown in Sec. III that, no matter
how CUTE or new physics affected the EFT chiral symmetry, the
chiral Ward identities and trace identities are consistent with
each other. On the other hand, our CUTE or new physics
interpretation of the anomalies also implies the requirements for
'building' models for CUTE or 'identifying' new physics: they must
yield the same EFT anomalies in the decoupling limit! In other
words, the anomalies in EFT are just the simplified
parametrization of the underlying structures, a generalization of
't Hooft's interpretation of chiral anomaly\cite{Hooft}.

In fact one can exploit the underlying theory scenario and the
decoupling behavior in the low energy regions beyond the
interpretations of anomalies. One could manipulate a simple proof
of the finiteness or non-renormalization of the Chern-Simons
action by employing gauge theory with heavy fermions as the
underlying theory, see Ref.\cite{Chern}.

In summary, we described a general parametrization of the
ill-defined Feynman loop amplitudes that is based on the existence
of a complete theory underlying the EFT's and investigated the
trace identities and chiral identities for certain two-point
functions. The anomalies are interpreted in the underlying theory
or new physics' perspective. As by-products, we showed that: (1)
trace identities are consistent with chiral identities no matter
what kind of regularization schemes are used: preserving chiral
symmetry or not; (2) some terms canonical in chiral identities are
anomalous in trace identities. Related remarks and speculations
were also presented.
\section*{Acknowledgement} The author is grateful to W. Zhu for
his continuing supports and helpful conversations on the topics
related to scaling and to Dr. Jian-Hong Ruan for her kind helps.
This work is supported in part by the National Natural Science
Foundation of China under Grant No.s 10075020 and 10205004, and by
the ECNU renovation fund for young researcher under No. 53200179.
\section*{Appendix A}
Dimensional regularization:
\begin{eqnarray}
I^{\epsilon}_a&=&\frac{i m^2}{(4\pi)^2}(\Gamma(\epsilon)+1-\ln
\frac{m^2}{4\pi \mu^2})=I^{\epsilon}_b,\\
I^{\epsilon}_{a^2}&=&\frac{i}{(4\pi)^2}(\Gamma(\epsilon)+1-\ln
\frac{m^2}{4\pi \mu^2}),\\
I^{\epsilon}_{ab}&=&\frac{i}{(4\pi)^2}(\Gamma(\epsilon)-\Delta_0),\\
I^{\epsilon}_{ab;\mu}&=&\frac{i p_{\mu}}{2(4pi)^2}
(\Gamma(\epsilon)-\Delta_0),\\
I^{\epsilon}_{ab;\mu\nu}&=&\frac{ig_{\mu\nu}}{2(4\pi)^2}(2p^2\int^1_0
dx (x^2-x)(\Gamma(\epsilon)+1-\ln \frac{D}{4\pi\mu^2})+
m^2(\Gamma(\epsilon)+1-\Delta_0))\nonumber \\
&&+\frac{ip_{\mu}p_{\nu}}{(4\pi)^2}\int^1_0 dx
(1-x)^2(\Gamma(\epsilon)-\ln \frac{D}{4\pi\mu^2}),\\
I_{a^2b}&=&\frac{-i}{2(4\pi)^2\Delta},\\
I_{a^2b;\mu}&=&\frac{ip_{\mu}}{(4\pi)^2p^2}(1-\frac{m^2}{\Delta}),\\
I^{\epsilon}_{a^2b;\mu\nu}&=&\frac{ip_{\mu}p_{\nu}}{2(4\pi)^2p^2}
(1-\frac{m^2}{\Delta})+
\frac{ig_{\mu\nu}}{4(4\pi)^2}(\Gamma(\epsilon)-\Delta_0),\\
D&=&m^2+p^2(x^2-x), \Delta_0=\int^1_0 dx \ln \frac{D}{4\pi\mu^2},
\frac{1}{\Delta}=\int^1_0 \frac{dx}{D}.
\end{eqnarray}The integrals $I_{a^2b}$ and $I_{a^2b;\mu}$ are
convergent.

Differential equation approach:
\begin{eqnarray}
I_a&=&-\frac{i }{(4\pi)^2}(m^2(C_0-1+\ln
\frac{m^2}{\mu^2})+C^{\prime})=I_b,\\
I_{a^2}&=&-\frac{i}{(4\pi)^2}(C_1+\ln
\frac{m^2}{\mu^2}),\\
I_{ab}&=&-\frac{i}{(4\pi)^2}(C_2+\Delta_0),\\
I_{ab;\mu}&=&-\frac{i p_{\mu}}{2(4\pi)^2}(C_3+\Delta_0),\\
I_{ab;\mu\nu}&=&-\frac{ig_{\mu\nu}}{2(4\pi)^2}(2p^2\int^1_0 dx
(x^2-x)(C_4-1+\ln  \frac{D}{\mu^2})+ m^2(C_5+\Delta_0))\nonumber \\
&& -\frac{ip_{\mu}p_{\nu}}{(4\pi)^2}\int^1_0 dx (1-x)^2(C_4+\ln
 \frac{D}{\mu^2})-\frac{ig_{\mu\nu}p^2}{4(4\pi)^2}C_{4;2}-
\frac{ig_{\mu\nu}}{2(4\pi)^2}C_5^{\prime},\\
I_{a^2b}&=&\frac{-i}{2(4\pi)^2\Delta},\\
I_{a^2b;\mu}&=&\frac{ip_{\mu}}{(4\pi)^2p^2}(1-\frac{m^2}{\Delta}),\\
I_{a^2b;\mu\nu}&=&\frac{ip_{\mu}p_{\nu}}{2(4\pi)^2p^2}
(1-\frac{m^2}{\Delta})-
\frac{ig_{\mu\nu}}{4(4\pi)^2}(C_6+\Delta_0),\\
I_{ab^2;\mu\nu}&=&-\frac{ip_{\mu}p_{\nu}}{2(4\pi)^2\Delta}
-\frac{3ip_{\mu}p_{\nu}}{p^2}
(1-\frac{m^2}{\Delta})-\frac{ig_{\mu\nu}}{4(4\pi)^2}(C_7+\Delta_0),\\
\Delta_0&=&\int^1_0 dx \ln \frac{D}{\mu^2}.
\end{eqnarray}In this approach there are 8 dimensionless arbitrary
constants ($C_0$, $C_1$, $C_2$, $C_3$, $C_4$, $C_5$, $C_6$, $C_7$
and $C_{4;2}$) and three dimensional arbitrary constants ($\mu,
C^{\prime}$ and $C_5^{\prime}$), all are mass independent. The
scale $\mu$ here should not be confused with the one in
dimensional regularization. However, there are natural relations
among these constants that will further reduce the number of
independent constants. For example, $I_a^2=\partial_{m^2}I_a,
I_{ab}|_{p=0}=I_{a^2}$, etc. Then we find that
$C_0=C_1=C_2=C_3=C_6=C_7=C_5+1=C_4+C_{4;2}$, hence there are at
most five independent constants. Note in employing these
relations, metric tensor contraction like $g_{\mu\nu}g^{\nu\mu}$
are not used as such operation is not well defined for the
ill-defined integrals, warning example can be found in dimensional
schemes.

\end{document}